\documentclass{article}
\usepackage{spconf,amsmath,graphicx}
\usepackage{multirow}
\usepackage{booktabs} %top rule
\usepackage{subcaption}
\let\OLDthebibliography\thebibliography
\renewcommand\thebibliography[1]{
    \OLDthebibliography{#1}
    \setlength{\parskip}{0pt}
    \setlength{\itemsep}{0pt plus 0.3ex}}
\usepackage{amsmath}
\DeclareMathOperator*{\argmax}{argmax}

\newcommand\correspondingauthor{\thanks{$^{*}$Corresponding author}}
\title{DEEPF0: END-TO-END FUNDAMENTAL FREQUENCY ESTIMATION FOR MUSIC AND SPEECH SIGNALS}
\name{Satwinder Singh \qquad Ruili Wang$^*$\correspondingauthor \qquad Yuanhang Qiu}
\address{School of Natural and Computational Sciences, Massey University, Auckland, New Zealand}
\begin{document}
\ninept
\maketitle
\begin{abstract}
We propose a novel pitch estimation technique called DeepF0, which leverages the available annotated data to directly learns from the raw audio in a data-driven manner. $f_0$ estimation is important in various speech processing and music information retrieval applications. Existing deep learning models for pitch estimations have relatively limited learning capabilities due to their shallow receptive field. The proposed model addresses this issue by extending the receptive field of a network by introducing the dilated convolutional blocks into the network. The dilation factor increases the network receptive field exponentially without increasing the parameters of the model exponentially. To make the training process more efficient and faster, DeepF0 is augmented with residual blocks with residual connections. Our empirical evaluation demonstrates that the proposed model outperforms the baselines in terms of raw pitch accuracy and raw chroma accuracy even using 77.4\% fewer network parameters. We also show that our model can capture reasonably well pitch estimation even under the various levels of accompaniment noise.        
\end{abstract}
\section{Introduction}
\label{sec:intro}
The fundamental frequency often represented by $f_0$ is the lowest and predominant frequency in a complex periodic signal. It is also referred to as the pitch of the waveform \cite{morrison2007ensemble}. However, there is a subtle difference between  $f_0$ and pitch since $f_0$ is perceived as the physical property of the audio signal, whereas pitch relates to the perceptual aspect of it. Nevertheless, outside the scope of psychoacoustics, both terms are used interchangeably in the literature \cite{gfeller2020spice} and in this paper as well. Pitch estimation has been studied for almost for the last five decades because of its central importance in various domains such as speech recognition \cite{ghahremani2014pitch}, speech synthesis \cite{reddy2020excitation},  and music information retrieval \cite{kum2019joint}. 
\par
There are many algorithms proposed in the past to carry out the task of pitch estimation. These algorithms can be categorized into two broad categories: digital signal processing (DSP) based approaches, and data-driven approaches. The signal processing based approaches can be further classified into the time-domain approaches  (RAPT \cite{talkin1995robust}, YIN \cite{de2002yin}, and pYIN \cite{mauch2014pyin}), frequency-domain approaches (SWIPE \cite{camacho2008sawtooth} and PEFAC \cite{gonzalez2014pefac}), or  hybrid (both time and frequency-domain) approaches (YAAPT \cite{kasi2002yet}). Most of them use a three-stage process consisting of preprocessing of a perceived signal (usually framing and signal conditioning) followed by possible candidate search using an auto-correlation function, cross-correlation function, or cepstrum function \cite{kim2018crepe}. Lastly, post-processing to track down the best possible candidates for $f_0$ using dynamic programming \cite{mauch2014pyin}. These approaches are computationally intensive,  not robust in noisy environments, fail when the pitch is rapidly changing, and do not learn anything from available data \cite{kim2018crepe}. On the other hand, data-driven approaches take full advantage of the available data and learn the estimation model based on the data itself. Most of the data-driven approaches are either based on traditional machine learning or deep learning approaches. Due to the availability of annotated pitch estimation datasets, and the success of deep learning models in various domains, it has become common practice to train the pitch estimation models in a data-driven manner. 
\begin{figure*}[!ht] % * for expanding figure to 2nd column
  \centering
  \includegraphics[width=16cm]{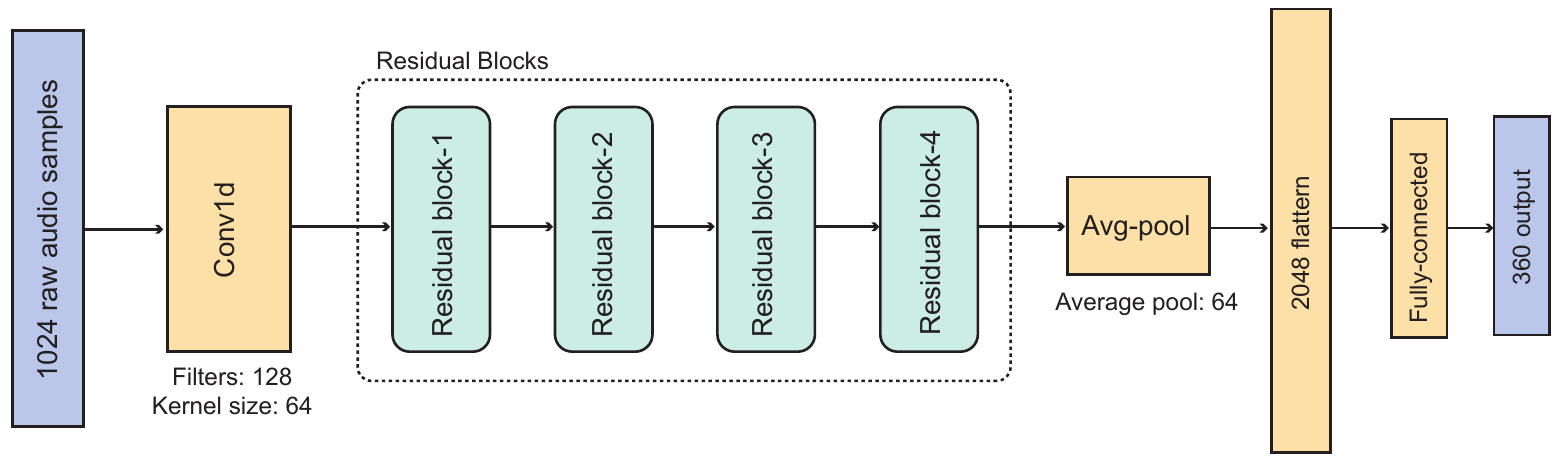} %\textwidth
  \caption{Network architecture of DeepF0.}
  \label{fig:architecture}
\end{figure*}
\begin{figure*}[!h] % * for expanding figure to 2nd column
  \centering
  \includegraphics[width=12cm]{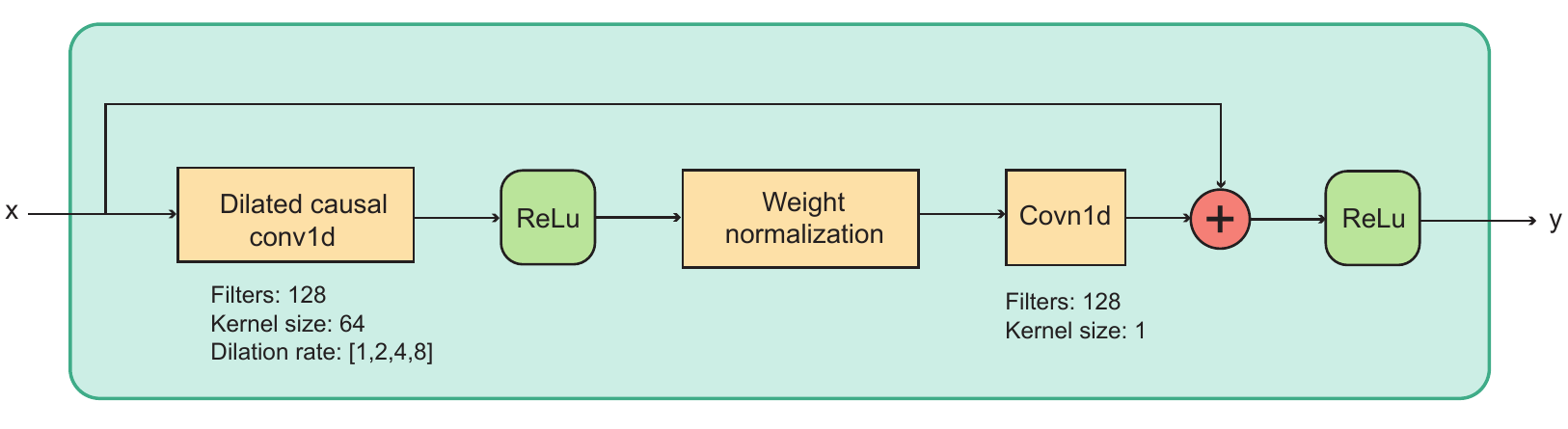}
  \caption{Internal view of a residual block of DeepF0.}
  \label{fig:residual-block}
\end{figure*}
Recently, numerous deep learning approaches were proposed either based on hand-crafted features or raw audio front-end. Many of the early work extracted hand-crafted features, which included constant-Q transform (CQT) \cite{gfeller2020spice} and spectral-domain features \cite{han2014neural}. While extracting the acoustic features from raw audio, there is always a chance of leaving out important features that might be crucial for pitch estimation \cite{dong2019vocal}. To deal with such a situation, many researchers attempted to exploit raw waveform as the front end features in various speech-related tasks \cite{dong2019vocal, sainath2015learning}. 
\par
In \cite{verma2016frequency}, a deep neural network (DNN) based pitch estimation model was proposed,  which operates on raw audio. Kim et al. \cite{kim2018crepe} designed a convolutional neural network (CNN) model that utilizes raw audio in the time domain and was able to outperform existing DSP based algorithms.  Similarly, Dong et al. \cite{dong2019vocal} proposed a convolutional residual network model to estimate pitch using raw polyphonic music. Even though these approaches perform better than DSP based algorithms, but these models still have very shallow receptive fields.  Authors in \cite{oord2016wavenet} and \cite{bai2018empirical} showed the effectiveness and applications of larger receptive in sequence modeling tasks. We intend to use that intuition in our pitch estimation task, where we can augment the network with dilation to have large memory or receptive field. We propose the DeepF0 model that is based on a dilated causal temporal convolutional network (TCN). Dilation in CNN increases the receptive field exponentially, without putting a computational burden on the network in terms of the number of network parameters used \cite{oord2016wavenet}. To stabilize the training of deep network, we introduce residual network blocks and skip connections to the network, which can increase training efficiency, and achieve high accuracy as well \cite{he2016deep}. The residual networks (also known as ResNet) have been successfully applied to a range of diverse areas of research \cite{oord2016wavenet, tian2019multi}.  
\par
We evaluated our proposed model on standard datasets that include MIR-1k \cite{hsu2009improvement}, MDB-stem-synth \cite{salamon2017analysis}, and PTDB-TUG \cite{pirker2011pitch}. These datasets contain audio samples of heterogeneous timbre and characteristics. We compared our approach with state-of-the-art CREPE and SWIPE baselines algorithms, where the former is deep learning based data-driven approach and the latter is a DSP based method. Empirical evaluation demonstrates that the proposed model yields state-of-the-art results in terms of pitch accuracy. Besides, the proposed method also outperforms the baselines in the presence of a reasonable amount of background noise. 
The rest of the paper is organized as follows: Section 2 outlines the proposed architecture. Section 3 describes the experimental setup. The results are discussed in Section 4, followed by Section 5, which concludes the paper.

\section{PROPOSED ARCHITECTURE}
The proposed approach is based on a dilated temporal convolution network. This type of architecture has been applied in the text-to-speech task (WaveNet \cite{oord2016wavenet}) but has not been applied in the pitch estimation task. The receptive field of the basic CNN is limited, which depends upon the linear depth of the network \cite{bai2018empirical}.  We can improve the receptive field by adding more convolutional layers, which will increase the receptive field linearly. However, this will put a computational burden on the network due to increased network parameters, and can also lead to a vanishing gradient problem. To address these issues dilated CNN is adopted \cite{oord2016wavenet}. In dilated convolutions (also referred to as convolutions with holes or atrous), we skip certain values to gain the receptive area which is usually larger than the filter size. This way we can achieve an exponentially large receptive field without even increasing the network parameters exponentially. 
 \par
The dilated convolutions are causal, which ensure that the current output is derived from the past outputs only and it does not look into future outputs. As we are increasing the depth of the network, which is ideal for learning robust representations, it can lead to the classical problem of vanishing gradients. Considering this issue, we adopt residual connections that resolve the problem of gradient vanishing by making new ways to flow the gradients \cite{oord2016wavenet}. It also makes sure that the higher layers perform as good as the deeper layers by learning through identity mapping while training a deeper network and is expressed as the following equation:

 \begin{equation}
 \label{eq:resnet}
   y = ReLu(x+\mathcal{F}(x))
\end{equation}
where $x$ is input, and $\mathcal{F}(x)$ represents a series of transformations like convolution operations and weight normalization.
\par
The DeepF0 model resamples the 16 kHz raw audio waveform in 1024 samples of audio frames, just like CREPE \cite{kim2018crepe} with a 360-dimensional output vector. The architecture of the model is shown in Figure \ref{fig:architecture}. The input is passed to 1D convolution with 128 filters and 64 kernel size. The big kernel size allows to have a wide receptive field and it also contributes to learn directly from the raw audio \cite{dong2019vocal}. The output of the first convolution goes through the residual blocks. Our residual block as shown in Figure \ref{fig:residual-block} consists of a 1D dilated causal convolution layer and a normal $1\times1$ convolutional layer followed by ReLu non-linearity \cite{nair2010rectified} and a weight normalization layer \cite{salimans2016weight}. To achieve extended receptive fields, we use dilation rate of $d=1,2,4,8$. In each residual block, we employ the residual/skip connections. 
The outcome of the last residual block is downsampled using average pooling with a pool size of 64 followed by a dense layer and sigmoid activation function.  The model uses a binary cross-entropy loss function to calculate the error between true $y_i$ and predicted values $\hat{y}_i$. The model is optimized using the Adam optimizer \cite{kingma2014adam} with a learning rate of 0.0002. Early stopping is enforced to ensure no overfitting when validation accuracy is not improving for 32 epochs. DeepF0 is trained using the Nvidia Geforce RTX 2080 Ti GPU.
\par
Following \cite{kim2018crepe}, the proposed model outputs 360-dimensional vector ($c_{1}-c_{360}$), which represents pitches on the logarithmic scale measured in terms of cents (a unit to measure small musical intervals). Each dimension of the output vector corresponds to the frequency bin that covers a frequency range from 32.70 Hz to 1975.5 Hz with 20 cents of intervals. The output vector estimates the Gaussian curve using the Gaussian kernel smoother \cite{kim2018crepe}. Afterwards, we calculate the pitch value by taking the local weighted average of pitches closest to frequency bins having the highest peak value as shown in Eq \ref{eq:local-average}. The resulted pitch values in cents are converted back to frequency equivalent (Hz) using Eq. \ref{eq:cents}, where $f$ is resulted frequency and 1200 being a single octave. $f_{ref}$ represents the reference frequency, which is set to 10 throughout our experimentation.  
\begin{equation}
   \hat c= \sum \limits_{i=m-4}^{m+4}(\hat y_i c_i)\bigg/ \sum \limits_{i=m-4}^{m+4}(\hat y_i)  \qquad
   m =  \argmax_i \hat y_i
    \label{eq:local-average}
\end{equation}

\begin{equation}
    f= f_{ref} \cdot 2^{\hat{c}/1200}
    \label{eq:cents}
\end{equation}

\section{EXPERIMENTAL SETUP}
\label{sec:pagestyle}

\subsection{Datasets}
The proposed model is trained and evaluated on three publicly available standard datasets, namely, MIR-1k \cite{hsu2009improvement}, MDB-stem-synth \cite{salamon2017analysis}, and PTDB-TUG \cite{pirker2011pitch}.  MIR-1k contains 1000 audio clips of people singing Chinese pop songs (11 males and 8 females) with pitch annotations. The right channel of the audio consists of the singing part, and the left channel holds musical accompaniment. The length of songs is between 4 and 13 seconds, which makes a total of 133 minutes of recordings. 
The MDB-stem-synth dataset contains 230 tracks resynthesized from the MedleyDB dataset \cite{bittner2014medleydb}. It consists of 418 minutes of diverse musical instruments and singing voices. Besides that, we also used the Pitch Tracking Database provided by the Graz University of Technology (PTDB-TUG). The dataset comprises of 4720 speech and laryngograph recordings of 20 native English speakers (10 males and 10 females) with a total length of 576 minutes. These three datasets have different characteristics, covering various musical instruments, singing voices, and speech signals.

\subsection{Methodology}
The proposed model is trained using 5-fold cross-validation with 60/20/20 split of train, validation, and test, respectively. The split is carried out in such a way that no artist/speaker/instrument overlaps with train and test splits. To investigate the model's robustness against background noise, we trained and evaluated the model on a dataset corrupted with musical accompaniment noise. We added a musical accompaniment noise to the original clean audio for the MIR-1k dataset at different signal-to-noise ratio (SNR) levels of 20dB, 10dB, and 0dB. % by audio degradation tool \cite{mauch2013audio}.

\subsection{Baselines}
We compare the proposed model with baseline models that include CREPE \cite{kim2018crepe} and SWIPE \cite{camacho2008sawtooth}. The DSP based SWIPE algorithm estimates pitch by template matching with the spectrum of the sawtooth waveform. On the other hand, CREPE is data-driven deep learning model, which is based on CNN and trained using multiple datasets (MIR-1k \cite{hsu2009improvement}, MDB-stem-synth \cite{salamon2017analysis},  MedlyDB \cite{bittner2014medleydb}, RWC-Synth \cite{mauch2014pyin},  Nsynth \cite{engel2017neural}, and Bach10 \cite{duan2010multiple}). We use the full version of the CREPE model (22.2M parameters) without Viterbi smoothing provided by the authors.  Both the chosen models directly operate on the raw waveform in the time domain. 

\subsection{Evaluation metrics}
To evaluate and compare the performance of the proposed model with the baselines, we use evaluation metrics defined in \cite{raffel2014mir_eval}. 
Raw Pitch Accuracy (RPA) measures the percentage of audio frames where the frequency estimate is accurate within the threshold value, which is 50 cents in our case. Raw Chroma Accuracy (RCA) also measures the percentage of audio frames where the frequency estimate is correct. However, the octave errors are ignored and mapped on to a single octave. Note that both RPA and RCA ignore voicing errors.

\section{Results and Discussion}
\subsection{Pitch Accuracy}
The proposed model is compared with state-of-the-art models that include CREPE \cite{kim2018crepe} and SWIPE \cite{camacho2008sawtooth}. The results are depicted in Table \ref{table:results}. Our proposed model outperforms the baseline models in terms of raw pitch/chroma accuracies on all the three datasets on clean audio. In terms of raw pitch accuracy, the DeepF0 model shows  1.33\% and 9.29\% of relative improvement compared with CREPE and SWIPE models on the MIR-1k dataset, respectively. A similar trend can be seen in raw chroma accuracy where the proposed model outperforms both the baseline models with no added noise across all the datasets. On the MDB-stem-synth dataset, DeepF0 achieves near-perfect pitch estimation with 98.38\% RPA and 98.44\% RCA in comparison with its baselines. We also evaluated our model on an additional dataset (PTDB-TUG), which has heterogeneous timbre (speaking voices) as compared to MIR-1k (singing voices) and MDB-stem-synth (musical instruments). Our model significantly performs better on PTDB-TUG in contrast to CREPE and SWIPE. On the PTDB-TUG dataset, CREPE performs worst of all the models. This could be attributed to the fact the model provided by the authors was not trained on the PTDB-TUG dataset. 
DeepF0 also shows more stability as it demonstrates consistently lower variance in pitch accuracy next to its baselines. 

\begin{table}[b]
\footnotesize
\centering
\caption{Average raw pitch accuracy and raw chroma accuracy and their standard deviation ($\pm$) tested on three different test datasets. }
\setlength{\tabcolsep}{1.8pt}
\label{table:results}
\renewcommand{\arraystretch}{1.2} % Default value: 1
\begin{tabular}{c|c|c|c|c|c}
\bottomrule
\multirow{2}{*}{Model} &\multirow{2}{*}{Params}  & \multirow{2}{*}{Metrics} & \multicolumn{3}{c}{Datasets}                     \\ \cline{4-6} 
                        &         &                 & MIR-1k          & MDB-stem-synth  & PTDB-TUG       \\ \toprule
\multirow{2}{*}{SWIPE} & \multirow{2}{*}{-} & RPA (\%)                 & 88.73 $\pm 5.43$          & 92.84 $\pm 9.59$         & 87.74  $\pm 7.17$        \\
                        && RCA (\%)                 & 89.24 $\pm 5.28$         & 93.83 $\pm 7.69$          & 88.93  $\pm 6.12$      \\ \hline
\multirow{2}{*}{CREPE} & \multirow{2}{*}{22.2M} & RPA (\%)                 & 96.51 $\pm 3.23$         & 97.22 $\pm 4.12$         & 78.18 $\pm 10.07$             \\
                       & & RCA (\%)                 & 96.84 $\pm 2.56$         & 97.55  $\pm 3.43$        & 79.81 $\pm 9.39$             \\ \hline
\multirow{2}{*}{DeepF0} & \multirow{2}{*}{5M} & RPA (\%)                 & \textbf{97.82 $\pm$ 3.34} & \textbf{98.38 $\pm$ 2.97} & \textbf{93.14 $\pm$ 3.32} \\
                       & & RCA (\%)                 & \textbf{98.28 $\pm$ 1.94} & \textbf{98.44 $\pm$ 2.87} & \textbf{93.47 $\pm$ 3.41} \\ \toprule
\end{tabular}
\end{table}

\begin{table}[!b]
\footnotesize
\centering

\caption{Average raw pitch accuracy and  raw chroma accuracy and their standard deviation ($\pm$) on the MIR-1k dataset with added noise on various levels of SNR.}
\setlength{\tabcolsep}{1pt}
\renewcommand{\arraystretch}{1.2} % Default value: 1
\label{table:noise}
\begin{tabular}{c|c|c|c|c|c}
\bottomrule
\multirow{2}{*}{Model}  & \multirow{2}{*}{Metrics} & \multicolumn{4}{c}{Noise Profile}                                \\ \cline{3-6} 
                        &                          & Clean          & 20dB           & 10dB          & 0dB            \\ \toprule
\multirow{2}{*}{SWIPE}  & RPA (\%)                 & 88.73 $\pm$ 5.43         & 84.45 $\pm$ 5.64          & 59.78 $\pm$ 11.58          & 32.04 $\pm$ 11.84        \\ 
                        & RCA (\%)                 & 89.24  $\pm$ 5.28      & 85.31 $\pm$ 5.19        & 62.85  $\pm$ 11.07       & 37.31 $\pm$  12.93       \\ \hline
\multirow{2}{*}{CREPE}  & RPA (\%)                 & 96.51 $\pm$ 3.23        & 96.49 $\pm$ 3.32        & \textbf{95.11 $\pm$ 4.65 } & \textbf{84.92 $\pm$ 10.70} \\ 
                        & RCA (\%)                 & 96.84 $\pm$ 2.56        & 96.96 $\pm$ 2.63        & 96.18 $\pm$ 3.35         & \textbf{87.85 $\pm$ 8.82} \\ \hline
\multirow{2}{*}{DeepF0} & RPA (\%)                 & \textbf{97.82 $\pm$ 3.34} & \textbf{97.39 $\pm$ 3.76} & 94.77$\pm$ 6.03 &    79.52$\pm$ 14.0         \\ 
                        & RCA (\%)                 & \textbf{98.28 $\pm$ 1.94} & \textbf{98.09 $\pm$ 2.10} & \textbf{96.35 $\pm$ 3.72} & 84.37 $\pm$ 10.71        \\ \toprule
\end{tabular}
\end{table}    

\subsection{Performance in noisy conditions}
Ideally, even in noisy environments, a model can still perform reasonably well. We put our proposed model into such testing scenarios by contaminating the signals with musical accompaniments at various levels of SNR on the MIR-1k dataset and results are presented in Table \ref{table:noise}.  In general, our proposed method achieves higher RPA and RCA as compared with the baselines under 10dB and 20dB noise. However, CREPE performs better when SNR is as low as 0dB. On the other hand, the performance of the SWIPE model is worst under 10dB and 0dB noise. Overall, we can say that DeepF0 achieves better performance under low to moderate noise scenarios and CREPE works better under moderate to high noise scenarios.

\begin{figure}[!t] % * for expanding figure to 2nd column
      \centering
      % include first image
      \includegraphics[width=8cm]{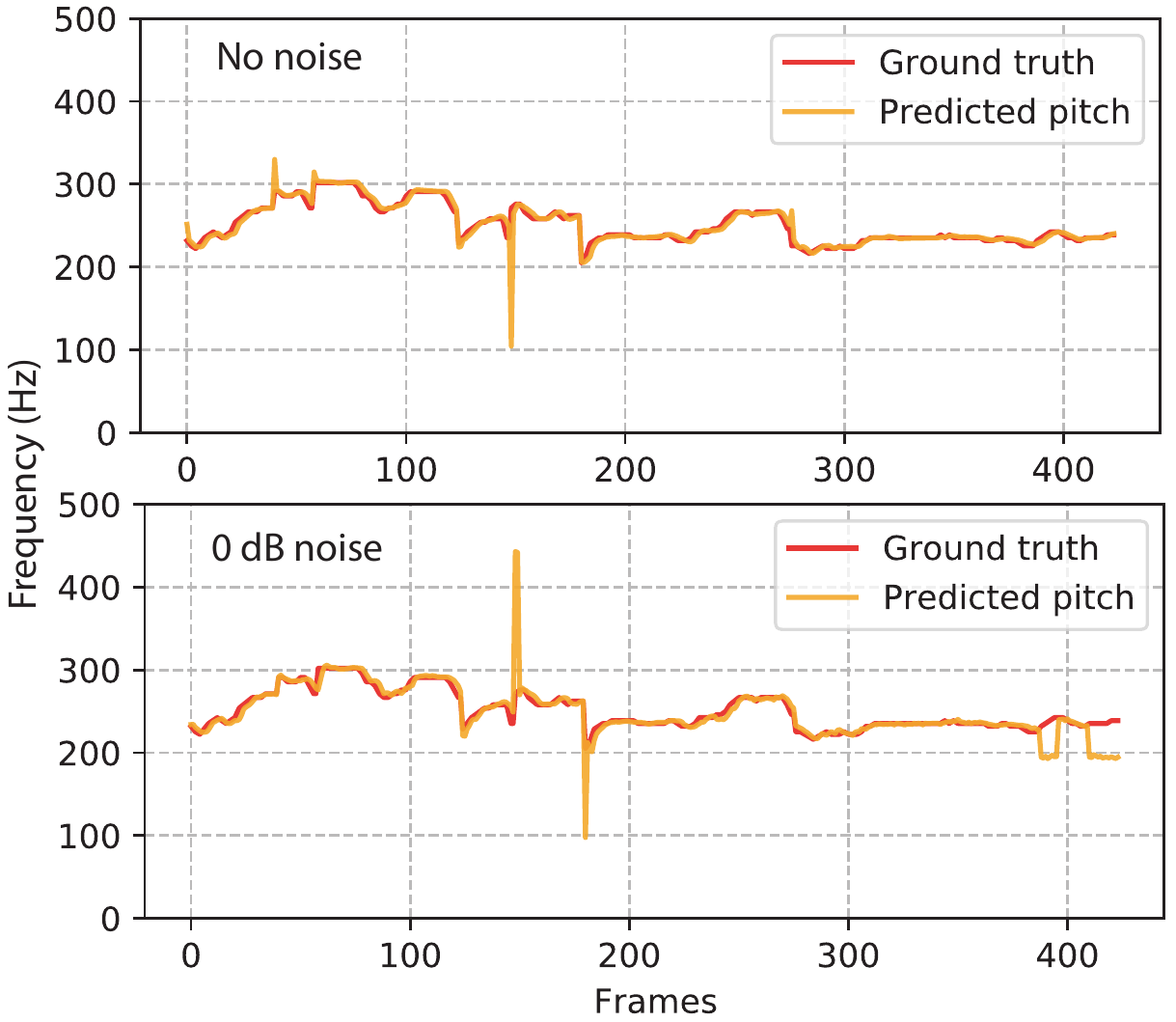}  
      \caption{The estimated pitch trajectories of DeepF0 in comparison with ground truth under clean (top) and 0dB noise (bottom). Under no noise scenario DeepF0 produces near perfect pitch estimation, while under noise there are few errors.}
      \label{fig:pitcch-plots}
\end{figure}

\begin{figure}[!b] % * for expanding figure to 2nd column
      \centering
      % include first image
      \includegraphics[width=8cm]{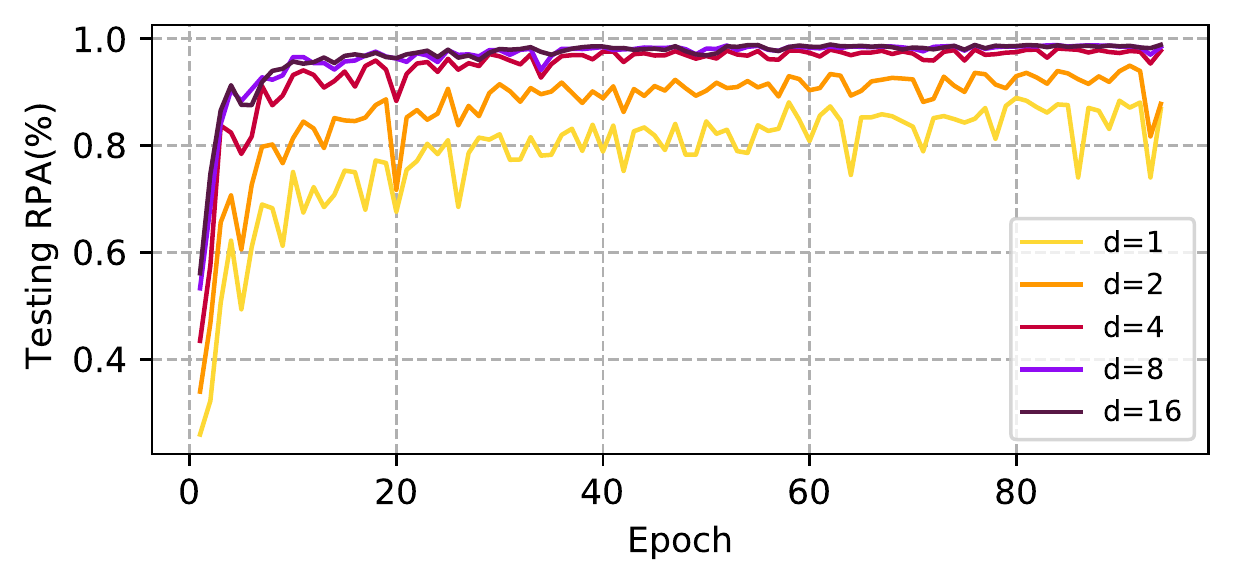}  
      \caption{Evaluation results of the proposed model with different dilation rates on the MDB-stem-synth dataset. Dilation rate $d=8$ shows the best results. }
      \label{fig:dilation-plots}
      %\vspace{-4mm}
\end{figure}
 
\subsection{Model analysis}
Our proposed model is more efficient in terms of the number of parameters used for training, which is around 5 million as compared with the CREPE model, which uses 22.2 million parameters. Thus, our DeepF0 model with 77.4\% fewer parameters can still outperform the CREPE model. Further, we analyze the role of the receptive field in the task of pitch estimation. We observe that a larger receptive field indeed improves the overall performance of the model. Our experiments are with dilation rate $d=1, 2, 4, 8, 16$ on the MDB-stem-synth dataset. The raw pitch accuracies are depicted in Figure \ref{fig:dilation-plots}. DeepF0 with $d=1$, which is basically a standard CNN, only achieves about 86.40\% of RPA and 86.55\% of RCA. Further $d=2,4$ improve the performance and able to achieve similar results as the CREPE model. However, the results are not very stable in terms of variance, which ranges from $\pm 6.85$ to  $\pm 18.11$ on dilation rate 1 to 4. We find that $d=8$ gives the best results in terms of RPA (98.38\% $\pm2.97$), RCA (98.44\% $\pm2.87$), and variance. DeepF0 does not show any performance improvements beyond the dilation rate of $d=8$. 
\par
We have also analyzed some of the design choices that we made while constructing the architecture of DeepF0 and the results of our ablation study are presented in Table \ref{table:ablation}. We find that residual connections are making a significant difference when it comes to stabilizing and speeding up the training process. Not only model converges fast, but it also helps in achieving higher performance with low variance. Besides this, we have not used dropout layers throughout our network as we observe that these layers seemed redundant in the presence of the weight normalization layer and have almost zero effect on the final results. 

\begin{table}[]
\centering\caption{Evaluation results of the ablation study of our DeepF0 model. Without residual connections accuracy of the model decrease. With the dropout layer included in the residual blocks, the performance more or less remains the same.}

\setlength{\tabcolsep}{9pt}
\renewcommand{\arraystretch}{1} % Default value: 1
\label{table:ablation}
\begin{tabular}{c|c|c}

\bottomrule
\multirow{2}{*}{Models}                   & \multirow{2}{*}{Metrics} & Dataset        \\\cline{3-3} 
                                          &                          & MIR-1k         \\ \toprule
\multirow{2}{*}{DeepF0 baseline}          & RPA(\%)                  & 97.82$\pm$3.34 \\
                                          & RCA(\%)                  & 98.28$\pm$1.94 \\ \hline
\multirow{2}{*}{w/o residual connections} & RPA(\%)                  & 97.54$\pm$3.61 \\ 
                                          & RCA(\%)                  & 97.89$\pm$2.42 \\ \hline
\multirow{2}{*}{w/ dropout}               & RPA(\%)                  & 97.83$\pm$3.28 \\
                                          & RCA(\%)                  & 98.24$\pm$2.18 \\ \toprule
\end{tabular}
%\vspace{-4mm}
\end{table}
\section{Conclusions}
In this paper, we propose a data-driven approach based on dilated temporal convolutional networks for the task of fundamental frequency estimation. The proposed DeepF0 model operates on a raw audio and outputs the pitch estimation. The experimental results performed on three heterogeneous datasets (singing voice vs speaking voices vs musical instruments) reveal that  DeepF0 outperforms existing baseline models like CREPE and SWIPE. Our proposed model does not only achieve better results but also used 77.4\% fewer parameters as compared with the CREPE model. Our model is also able to perform reasonably well under low to moderate noise. Further, we gain crucial insights about the large receptive field, which was not there in earlier models. We find that the length of the receptive field of the network is very significant in pitch estimation, which aids in achieving excellent results with consistently low variance. 
In the future, we would like to improve the noise robustness of our proposed model by introducing changes in architectural design, data augmentation and speech enhancement techniques \cite{qiu2020adversarial}. The performance can be further improved by post-processing the pitch estimate through temporal smoothing techniques.
\clearpage
\bibliographystyle{IEEEbib}
\bibliography{strings,refs}

\end{document}